\documentclass[aps,prb,twocolumn,superscriptaddress,showpacs]{revtex4}
\usepackage{graphicx}
\usepackage{multirow}
\usepackage[usenames]{color}
\usepackage{array}

\begin{document}

\title{On the half-metallicity of Co$_2$FeSi Heusler alloy: an experimental and {\bfseries \slshape ab initio} study}

\author{L. Makinistian}
\affiliation{INTEC-CONICET, G\"uemes 3450, 3000 Santa Fe, Argentina}

\affiliation{Facultad de Ingenier\'ia, Universidad Nacional de Entre
R\'ios, 3101 Oro Verde, Argentina}

\author{$\text{Muhammad M.}$ Faiz}\author{Raghava P. Panguluri}
\affiliation{Department of Physics and Astronomy, Wayne State\ University,
Detroit MI, 48201}

\author{B.Balke}\author{S.Wurmehl}\author{C. Felser}
\affiliation{Institut f\"ur Anorganische Chemie und Analytische Chemie, Johannes
Gutenberg-Universit\"at, D-55099 Mainz, Germany}

\author{E. A. Albanesi}
\affiliation{INTEC-CONICET, G\"uemes 3450, 3000 Santa Fe, Argentina}

\affiliation{Facultad de Ingenier\'ia, Universidad Nacional de Entre
R\'ios, 3101 Oro Verde, Argentina}
\author{A. G. Petukhov}
\affiliation{South Dakota School of Mines, Department of Physics, Rapid
City, South Dakota 57701-3995}

\author{B. Nadgorny}
\affiliation{Department of Physics and Astronomy, Wayne State\ University,
Detroit MI, 48201}

\begin{abstract}
Co$_2$FeSi, a Heusler alloy with the highest magnetic moment per unit cell and the highest Curie temperature, has largely been described theoretically as a half-metal. This conclusion, however, disagrees with Point Contact Andreev Reflection (PCAR) spectroscopy measurements, which give much lower values of spin polarization, $P$. Here, we present the spin polarization measurements of Co$_2$FeSi by the PCAR technique, along with a thorough computational exploration, within the DFT and a GGA+U approach, of the Coulomb exchange U-parameters for Co and Fe atoms, taking into account spin-orbit coupling. We find that the orbital contribution ($m_o$) to the total magnetic moment ($m_T$) is significant, since it is at least $3$ times greater than the experimental uncertainty of $m_T$. Account of $m_o$ radically affects the acceptable values of U. Specifically, we find no values of U that would simultaneously satisfy the experimental values of the magnetic moment and result in the half-metallicity of Co$_2$FeSi. On the other hand, the ranges of U that we report as acceptable are compatible with spin polarization measurements (ours and the ones found in the literature), which all are within approximately $40$-$60$\% range. Thus, based on reconciling experimental and computational results, we  conclude that: a) spin-orbit coupling cannot be neglected in calculating Co$_2$FeSi magnetic properties, and b) Co$_2$FeSi Heusler alloy is not half-metallic. We believe that our approach can be applied to other Heusler alloys  such as Co$_2$FeAl.
\end{abstract}
\pacs{71.20.-b, 71.20.Be, 75.70.Cn}

\maketitle

While Heusler compounds have been known for more than a hundred years \cite{Heusler03}, they drew a remarkable amount of attention~\cite{Kubler83,Galanakis02,Fecher06, Miura04,Hamaya2012} ever since the prediction by de Groot \emph{et al.}\cite{Groot83,Groot84} in the early 1980's that some of these alloys would have a metallic band structure for the majority spin channel and a semiconducting band structure for the minority one, resulting in 100\% spin-polarization ($P$) at the Fermi level. Such {\sl half}-metallic (HM) 
materials, with high values of $P$ and Curie temperature ($T_c$), are  excellent candidates for  spintronic applications~\cite{Zutic04,Felser07} (e.g., magnetic random access memories (MRAM)~\cite{Parkin99} 
utilizing the giant magneto-resistance spin-valve effect \cite{Dieny91}
in magnetic tunnel junctions \cite{Moodera95} ).
Specifically for Co$_{\text{2}}$FeSi, high and low temperature magnetometry experiments~\cite{Wurmehl05}  showed that it is a 
Heusler compound with the highest magnetic moment ($(5.97\pm 0.05)\mu_{B}$ per unit cell, at $5$ K) and the highest 
$T_c$ ($1100\ K$) among other Heusler alloys. 

Detailed computational studies have indicated that 
an orbital-dependent potential accounting for a moderate Coulomb-exchange interaction must be included in self-consistent calculations to 
simultaneously 
replicate both the experimental equilibrium lattice parameter ($5.64\ \text{\AA}$) and the measured magnetic moment of Co$_{\text{2}}$FeSi alloy.\cite{Wurmehl05,Kandpal06}
These studies have also 
revealed that a total spin magnetic moment 
$\sim 6\ \!\mu_B$ 
can be obtained only for the effective Coulomb-exchange interaction parameters~\cite{Anisimov97} ($U_{eff}=U-J$) 
falling within the ranges of $2.5$-$5\ eV$ and $2.4$-$4.8\ eV$ for the d-orbitals of Co and Fe atoms, respectively.
Even though Co$_2$FeSi alloy appears to be half-metallic only theoretically and only under stringent conditions
on $U_{eff}$, it  has been extensively referenced in the literature as such.~\cite{Gercsi2006,Gercsi2007,Takamura2008,Yamada2010,Kasahara2010,Yamada2011b}
 This prediction, however, is 
 at odds with several experimental measurements based on Point Contact Andreev Reflection (PCAR) spectroscopy~\cite{Gercsi2006,Karthik2007,Yamada2011b}, which yield values of $P\sim\!\! 50\%$, far lower than $100\%$.

The goal of this Rapid Communication is to reconcile the results of computational predictions with experimental measurements for Co$_2$FeSi. First, we present the results of our own PCAR measurements of $P$ and compare them with those available in the literature. Second, we perform a thorough computational exploration of the Coulomb exchange $U$-parameter space for the 3d-orbitals of Co and Fe atoms in Co$_2$FeSi, seeking the 
domain of parameters allowing to replicate 
the experimental measurements.

The samples were prepared by arc melting of stoichiometric amounts of the constituents in an argon atmosphere at $10^{^{-4}}$ mbar. The polycrystalline ingots were then annealed in an evacuated quartz tube at 1273 K for 21 days. This procedure resulted in samples exhibiting the Heusler type L2$_{\text{1}}$ structure, which was verified by X-ray powder diffraction (XRD) using excitation by Mo K$_{\alpha1}$ radiation. Flat disks were cut from the ingots and polished for spectroscopic investigations of bulk samples. X-ray photo emission (ESCA) was used to verify the composition and to check the cleanliness of the samples. After removal of the native oxide from the polished surfaces by Ar$^{\text{+}}$ ion bombardment, no impurities were detected with ESCA. 

\begin{figure}[ht]
\includegraphics[width=87mm]{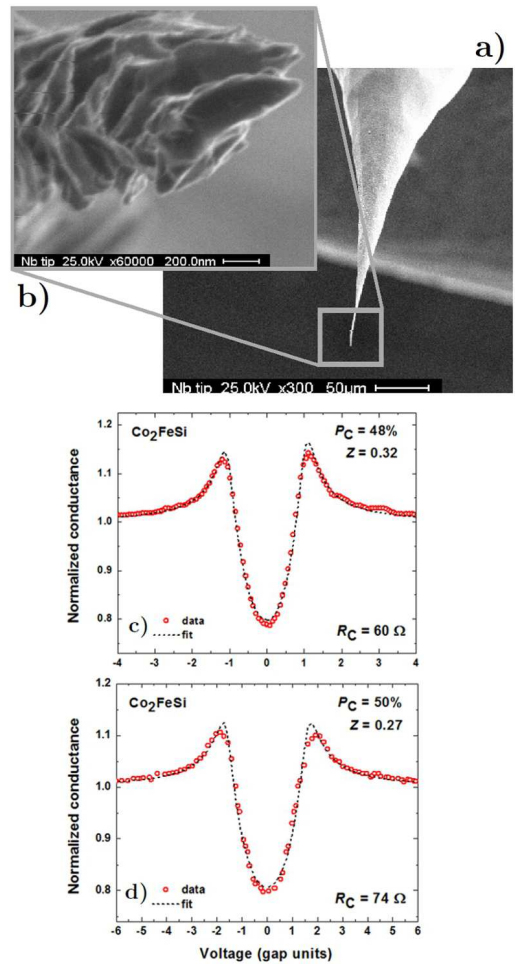} 
\caption{(Color online) Superconducting Nb tip: a) shows a scanning electron micrograph
of a tip ($\times 300$), and b) shows just the apex of a tip from a different angle and a greater amplification ($\times 60000$). c-d) Optimum normalized conductance curves for Nb/Co$_2$FeSi contacts as a function of voltage with Nb superconducting gap, $\Delta\!=\!1.5$ meV and $T\!=\!1.2$ K. The red circles represent the experimental data and the dashed black line is the fit.}
\label{fpcar} 
\end{figure}

For the PCAR measurements, niobium superconducting tips are fabricated by electrochemical etching of 250 $\mu$m thick niobium wire in a solution of HNO$_{3}$, HF, and CH$_{3}$COOH, with a mixing ratio of 5:4:1 by volume. The wire was kept at a positive potential with respect to the graphite counter electrode. The applied voltage was optimized at $\sim 8\!-\!10$ V for the output current of $\sim 800$-0 mA to get a sharp tip, see Fig. \ref{fpcar}a) and b). Just before the measurements the tip was briefly dipped into the HF solution. A freshly etched superconducting Nb tip (bulk T$_{C}$ $\sim 9.3$ K) was then mounted onto a shaft connected to a differential type screw that could be driven manually by $10$ $\mu$m per revolution. For the low temperature measurements both the tip and the sample were immersed into a liquid He bath. The current-voltage ($I\!-\!V$) measurements were taken using a four-probe technique, with the differential conductance $dI/dV$ obtained by standard ac lock-in detection at a frequency of $2$ kHz within the temperature range $1.2\!-\!4.2$ K. Typical results for normalized conductance $G(V)/Gn$ as a function of voltage V are shown in Fig. \ref{fpcar}c). At least $15$ different junctions with the contact resistance $1\ \Omega\le R_C \le 100 \Omega$ were measured and analized. To extract the values of spin polarization, $P$, the conductance curves for each junction were fitted with the modified\cite{Mazin2001} BTK model~\cite{Blonder82}, using the value of Nb superconducting gap, $\Delta\!=\!1.5\ meV$. As the values of $P$ were found practically independent of the interface transparency Z, the values for individual contacts were averaged out. The red circles in Fig. \ref{fpcar}c) represent the experimental data and the dashed black line is the fit. We find that the average value of spin polarization $\langle P\rangle\!=\!48\pm 3 \%$, Table \ref{tablePs}, shows that our results are in good agreement with the other PCAR measurements available in the literature.

\begin{table}[ht]
\caption{Experimental spin polarizations of Co$_2$FeSi by the PCAR technique.\label{tablePs}}
\begin{ruledtabular}
\begin{tabular}{ccc}
 P(\%) & PCAR on Co$_2$FeSi & Ref.\\ \hline
 $48\pm 3$ & bulk @$1.2\ K$ & this work\\
 $49\pm 2$ & thin films on MgO (001) @$4.2\ K$ & ~\cite{Gercsi2006}\\ 
 $57\pm 1$ & bulk @$4.2\ K$& ~\cite{Karthik2007}\\
 $59\pm 2$ & thin films on n-Ge(111) @$4.2\ K$& ~\cite{Yamada2011b}\\
\end{tabular}
\end{ruledtabular}
\end{table} 

\begin{figure*}[!]
\includegraphics[width=178mm]{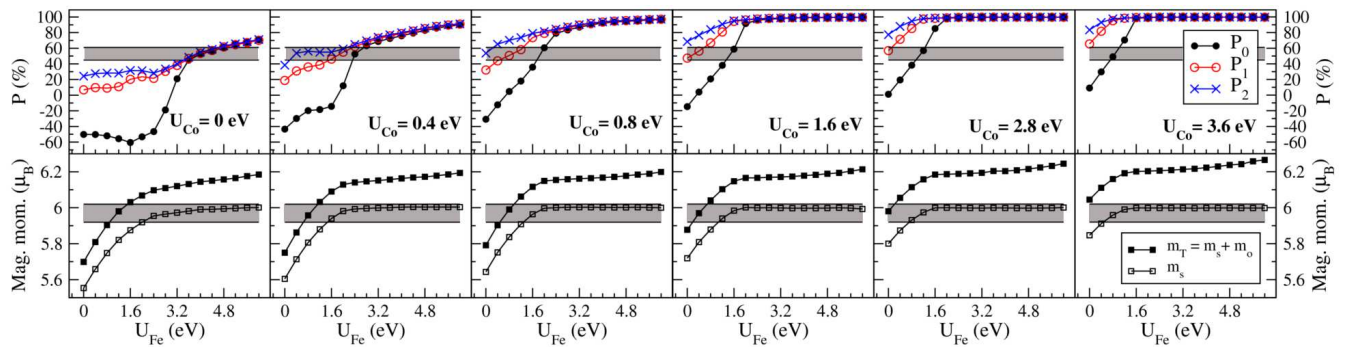} 
\caption{(Color online) Spin polarizations ($P_0$, $P_1$ and $P_2$, see text for details), and spin ($m_s$) and total (spin plus orbital: $m_T=m_s+m_o$) magnetic moment for selected values of $U_{Co}$ (from left to right: $0$, $0.4$, $0.8$, $1.6$, $2.8$, $3.6\ eV$) and $U_{Fe}$ in the range $0$-$6\ eV$. The gray shaded areas mark the experimental ranges for total magnetic moment~\cite{Wurmehl05}: $m_T=5.97\pm 0.05$ and the spin polarization (see Table \ref{tablePs}): $45<P<61$.}
\label{fexploringUspace} 
\end{figure*}

In order to proceed, we need to calculate the spin polarization, $P_{n}$, that can be defined by~\cite{Mazin1999,Nadgorny01,Sheet05}:
\begin{equation}
P_{n}=\frac{\langle N_{\uparrow}v_{\uparrow}^{n}\rangle -\langle N_{\downarrow}v_{\downarrow}^{n}\rangle}{\langle N_{\uparrow}v_{\uparrow}^{n}\rangle +\langle N_{\downarrow}v_{\downarrow}^{n}\rangle}\times 100\label{eqp}
\end{equation}
where the averages are taken upon all the sheets of the Fermi surface, and the exponent $n$ depends on the details of the experimental technique. For spin-resolved photoemission $n\!=\!0$ ($P_0$ is the ``static'' or ``intrinsic'' spin polarization); $n\!=\!1$ corresponds to experiments in the ballistic transport regime, whereas $n\!=\!2$ describes experiments dominated by diffusive transport. Ideally the PCAR experiments are done in the ballistic (Sharvin) regime, but if the mean free path is smaller than the minimum size of the contact, they can only be performed in the diffusive regime.~\cite{Soulen98,Bahramy07} It is clear from Eq.~(\ref{eqp}) that while $P_0$ can be directly calculated from the spin polarized density of states (DOS), $P_1$ and $P_2$, in addition to the DOS, also require the respective Fermi velocities. Following the approach of Scheidemantel {\sl et al.}~\cite{Scheidemantel03}, we calculated the Fermi velocity directly from the matrix elements of the momentum operator (instead of differentiating Bloch energies $E_{i,\vec{k}}$ with respect to $\vec{k}$) according to:
\begin{equation}
\vec{v}_{i,\vec{k}}=\frac{1}{m}\langle\psi_{i,\vec{k}}|\ \hat{\vec{p}}\ |\psi_{i,\vec{k}}\rangle
\end{equation}
where $i$ is the band index. These matrix elements can be readily generated  using the optical package of the full potential linearized augmented plane waves (FP-LAPW) \cite{Koelling75,Andersen75,Jepsen82,Weinert82,Jansen84,Mattheiss86} WIEN2k code\cite{Schwarz02},  and further used as an input in Eq.~(\ref{eqp}) to calculate $P_1$ and $P_2$.  We tested this procedure for pure bcc Fe and fcc Ni.  As we can see from Table \ref{tableNiFe} our results are in satisfactory agreement with other calculations and experimental data available in the literature.
\begin{table}[h]
\caption{Our calculated spin polarizations (in {\%}) for bcc Fe and fcc Ni, against others' calculations (figs. $2$-$3$ in $^a$Ref.~\cite{Bahramy07}, $^b$Ref.~\cite{Nadgorny2000}) and experiments ($^b$Ref.~\cite{Nadgorny2000}, $^c$Ref.~\cite{Soulen98}).}\label{tableNiFe}
\begin{ruledtabular}
\begin{tabular}{ccccccc}
 & \multicolumn{3}{c}{bcc Fe} &  \multicolumn{3}{c}{fcc Ni}\\ \cline{2-4} \cline{5-7}
  & Ours & Others &Exp.&  Ours & Others&Exp.\\ \hline
$P_0$  & $51$ & $58^a$, $59^b$& - & $-80$ & $-81^a$& - \\
$P_1$  & $39$ & $39^a$, $33^b$& - &$-45$ & $-48^a$& - \\
$P_2$  & $37$ & $33^a$, $21^b$& - & $5$ & $2^a$& - \\
$P_{exp}$ & -& -& $44\pm 3^b$, $40$-$48^c$& -&  -& $40$-$47.5^c$
\end{tabular}
\end{ruledtabular}
\end{table}

We carried out all our calculations with the  WIEN2K code, using the generalized gradient approximation (GGA) in the formal parameterization scheme of Perdew-Burke-Ernzerhof (PBE)~\cite{Perdew96} and the experimental lattice parameter of $5.64$ {\AA} for the cubic Co$_2$FeSi (crystallographic details of its structure can be found in Ref.~\cite{Aniruddha01}). Muffin tin radii ($R_{MT}$) of $2.32$, $2.32$, and $2.18$ atomic units were used for the Co, Fe, and Si atoms, respectively. The $R_{MT}^{^{*}{}}\cdot K_{max}$ product, where $R_{MT}^{^{*}}$ is the smallest of all muffin tin radii, and $K_{max}$ is the plane wave cut-off, was set equal to $7$ (implying a plane wave expansion cut-off of $\sim9.10$ Ry), and the energy threshold between core and valence states used was of $-6.0$ Ry. Integration in the irreducible Brillouin zone was carried out over $641$ k-points and convergence was set to simultaneously be better than $10^{-4}$ (Ry) for the total energy and $10^{-3}$ (au) for the total electronic charge. Spin-orbit coupling was included in all calculations in order to obtain not only the total spin, but also the total orbital moment. Within the LDA(GGA)+U scheme~\cite{Anisimov97} we have modified the 3d orbitals of Co and Fe atoms, and explored a square mesh of $16\!\times\!16$ points in the $U_{Co}$-$U_{Fe}$ space (resulting in a total of $256$ self-consistent calculations), with $0\leq U_{Co}\leq6\ eV$ and $0\leq U_{Fe}\leq6\ eV$. Since it has been shown~\cite{Petukhov2003} that the choice of different ``flavors'' of the double counting correction of the +U method can be critical, we tested two of them: one, the so-called ``SIC'' (self-interaction correction), introduced by Anisimov et al.~\cite{Anisimov1993,Anisimov97}, and the ``AMF'' (around mean field), introduced by Czy$\dot{\text{z}}$yk and Sawatzk~\cite{Czyzyk1994}. The latter yielded the total magnetic moments (spin plus orbital terms) lower than the lowest bound of the experiments ($5.92\ \mu_B$)~\cite{Wurmehl05} up to $U\sim 5\ eV$, so we found it inadequate for modeling Co$_2$FeSi. Thus, all the results reported here were obtained using the SIC flavor of the method.

\begin{figure}[ht!]
\includegraphics[width=85mm]{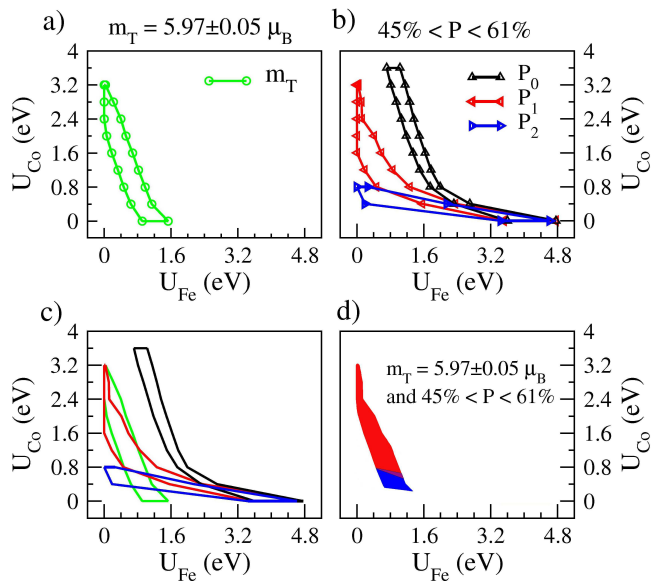} 
\caption{(Color online)  The area within the closed loops correspond to values of U for which a) the calculated total magnetic moment, $m_T$, agrees with experiment; b) the spin polarizations ($P_0$, $P_1$, or $P_2$) are within the experimental PCAR values. c) The four loops of a) and b) are superimposed (omitting the symbols) to aid the eye. d) Intersection of the areas of a) and b), i.e., the values of U for which both calculated $m_T$ and $P$ values fall within experimental values.}
\label{fUranges} 
\end{figure}

Representative sampling of our results is shown in Fig. \ref{fexploringUspace}, where, the total spin moment ($m_s$), the total magnetic moment ($m_T$, i.e., total spin ($m_s$) plus total orbital ($m_o$) moment), and the spin polarizations $P_0$, $P_1$, $P_2$  are presented as a function of $U_{Co}$ and $U_{Fe}$. It is clear that for a system where spin-orbit coupling is important, such as the case for Co$_2$FeSi, the spin-orbit interaction induces a strong orbital component in the $m_T$, which turns out to be significant and cannot be neglected: it shifts to lower energies, and simultaneously narrows down, the range of $U_{Fe}$ that yields a total magnetization within the experimental margin of error, $m_T=5.97\pm 0.05$~\cite{Wurmehl05} (see the gray shaded area in the bottom panels of Fig. \ref{fexploringUspace}). One can also see from Fig. \ref{fexploringUspace} that for $U_{Co}\geq 3.6\ eV$ none of the $U_{Fe}$-values would yield the value of $m_T$ within the experimental range. The total spin moment is also shown in all of the bottom panels (it is fixed at $6\ \mu_B$ beyond a certain value of $U_{Fe}$). The analysis of all our calculations allows us to produce Figs. \ref{fUranges}a) and b), that show the values of the U-parameter (areas within the closed loops in the figure) for which the calculated total magnetic moment and spin polarizations ($P_0$, $P_1$, or $P_2$) are within the experimental range. Finally, Fig. \ref{fUranges}d) shows the values of $U$ for which our calculations simultaneously agree with the results of both: magnetometry and PCAR spectroscopy (the latter is taken over the entire range, $45\%<P<61\%$, defined by our measurements and others, see Table \ref{tablePs}). Except for the lower area ($U\sim 0.4\ eV$, which corresponds to $P_2$), all other values correspond to $P_1$, i.e., the expected ballistic transport between the superconducting tip and the sample in the PCAR experiments with Co$_2$FeSi. Thus, in view of our findings, the results of magnetometry measurements indicate that Co$_2$FeSi is actually not a half-metal, and the determination of $P$ using PCAR spectroscopy is fully compatible with this prediction.

Chalsani {\sl et al.}~\cite{Chalsani2007} mentioned that the results of the PCAR technique can be affected by a number of factors, such as the geometry of the contact and interactions between the sample and the tip through surface states; however, {\sl ab initio} calculations by Khosravizadeh {et al.}~\cite{Khosravizadeh2009} showed that the surface of Co$_2$FeSi is also non half-metallic, and the loss half-metallicity of Co$_2$FeSi, which we have found, cannot be attributed to the surface effects. Moreover, our results for a perfect bulk material would persist, regardless of whether or not anti-site defects, finite temperature~\cite{Attema2004}, and crystallographic disorder~\cite{Orgassa1999,Panguluri2009}, could induce additional loss of half-metallicity by the appearance of spin states in the minority spin channel. The orbital moments we calculated in Co$_2$FeSi depend on the U-parameter, ranging from $0.145\ \mu_B$ for $U_{Co} = U_{Fe} = 0$, to $0.482\ \mu_B$ for $U_{Co} = U_{Fe} = 6\ eV$. One key idea presented here is that even the lowest orbital moment obtained ($m_o =0.145\ mu_B$) is almost three times greater than the experimental uncertainty ($0.05\ \mu_B$) in the magnetometry measurements of the total magnetic moment per formula unit~\cite{Wurmehl05} ($m_T = 5.97\pm 0.05 \mu_B$), as Fig. \ref{fexploringUspace} clearly demonstrates. In addition, the total spin moment as is shown in Fig. \ref{fexploringUspace} (in all the bottom panels) is fixed at $6\ \mu_B$ beyond a certain value of $U_{Fe}$. It is also seen that for $U_{Co} = 3.6\ eV$ there is no $U_{Fe}$ value that would yield an $m_T$ within the experimental range. Our calculations of $m_o$, are also supported by the systematic study by Galanakis~\cite{Galanakis2005}, performed on nine full-Heuslers alloys, and while the author does not study Co$_2$FeSi, he presents the results on similar compounds, Co$_2$FeAl and Co$_2$MnSn, citing for the former the greatest total orbital moment calculated in that work to be $0.149\ \mu_B$, which is in excellent agreement with our $0.145\ \mu_B$ value. 

In summary, our calculations provide a strong evidence that the orbital component of the total magnetic moment in Co$_2$FeSi cannot be neglected. By taking it into account we identify the ranges of the $U$-parameters compatible with both the magnetometry and PCAR (ours and others) measurements. Based on the range of the U-parameters, we conclude that Co$_2$FeSi is not a half-metal. We believe that our approach will be applicable to other compounds similar to Co$_2$FeSi, such as Co$_2$FeAl and Co$_2$MnSn. 

This work was supported by DOE Grant No. DE-SC0004890 at SDSM\&T, the Consejo Nacional de Investigaciones Cient\'ificas y T\'ecnicas (CONICET), and the Universidad Nacional de Entre R\'ios (UNER), Argentina.


\end{document}